\documentclass[prl,showpacs,showkeys,twocolumn,10pt]{revtex4-1}
\usepackage{amsfonts,bbold,amsmath,amssymb,graphicx,epstopdf,verbatim,dsfont,color}
\usepackage[english]{babel}

\begin{document}

\title{Information propagation and equilibration in long-range Kitaev chains}
\author{Mathias Van Regemortel}
\email{Mathias.VanRegemortel@uantwerpen.be}
\affiliation{TQC, Universiteit Antwerpen, B-2610 Antwerpen, Belgium}
\author{Dries Sels}
\affiliation{TQC, Universiteit Antwerpen, B-2610 Antwerpen, Belgium}
\affiliation{Department of Physics, Boston University, Boston, MA 02215, USA}
\author{Michiel Wouters}
\affiliation{TQC, Universiteit Antwerpen, B-2610 Antwerpen, Belgium}
\date{\today}

\begin{abstract}
We study the propagation of information through a Kitaev chain with long-range pairing interactions. Although the Lieb-Robinson bound is violated in the strict sense for long-range interacting systems, we illustrate that a major amount of information in this model still propagates ballistically on a light cone. 
We find a pronounced effect of the interaction range on the decay of the mutual information between spatially disconnected subsystems. A significant amount of information is shared at time-like separations. This regime is accompanied by very slow equilibration of local observables. As the Kitaev model is quasi-free, we illustrate how the distribution of quasi-particle group velocities explains the physics of this system qualitatively.
\end{abstract}
\maketitle

Much of our understanding of the non-equilibrium dynamics of locally interacting closed quantum system is deduced from  Lieb-Robinson bounds~\cite{LR}. Even without imposing Lorenz invariance this bound shows that the effect of a local perturbation can not be measured elsewhere in the system outside an effective causal cone. The emergent causality puts severe constraints on the dynamical behaviour of the system. Not only does the Lieb-Robinson bound provide an intuitive picture of the spreading of correlations in the system, it has also enabled numerous proofs on the distribution of correlations and entanglement \cite{Bravyi,Eisert06}, as well as on equilibrium properties \cite{Hastings,Nachtergaele} of condensed matter systems.

The perfect isolation of a quantum system from its environment has never been approached so closely as in experiments with cold atoms and ions. They offer a versatile platform to study the non-equilibrium behaviour of many-body systems \cite{Polkovnikov}, such as Lieb-Robinson bounds and the lightcone-like spreading of correlations~\cite{Cheneau}.  Due to recent advances in cold atoms and trapped ions experiments it has now become possible to study also the behaviour of systems with long and variable-range interactions \cite{Richerme,Britton,Jurcevic}. When the interactions become long-range the system correlations do not need to obey the Lieb-Robinson bound. Until recently very little was know about the behaviour of those systems as analytical results are scarce and known bounds~\cite{Hastings} for long-range interactions were too loose to provide any insight. There has been considerable theoretical progress since then~\cite{Schachenmayer,Hauke,Gong,Storch,vdWorm,nedzha}.

Thanks to its integrability, the Ising chain in a transverse field is the paradigmatic model for studying the dynamics of information propagation in short range interacting systems \cite{Calabrese}. Unfortunately a long range interaction breaks integrability and full numerical simulations are required \cite{Schachenmayer,Hauke}. Integrable models with long range interactions that have been considered, consist of free bosons and fermions with long-range hopping ~\cite{nedzha,Storch}, and also the long-range Kitaev (LRK) model~\cite{Vodola}. It is the extension of the short-range Kitaev chain \cite{Kitaev} with pairing interactions that decay as $1/r^\alpha$. 
The interaction range affects the entanglement in the ground state, violating the  area law, as well as the entanglement dynamics after a quench.

The near perfect isolation of modern many-body systems has also renewed interest in the fundamental physics of thermalization in closed quantum systems \cite{Polkovnikov,EisertNP}. In the case of integrable models, it has become clear since the pioneering experiments by Kinoshita {\em et al.} \cite{Kinoshita} that a generalized Gibbs ensemble (GGE) \cite{Rigol} is required for the description of the long-time equilibrated state. 

In this letter, we will study the issues of information propagation and equilibration in the LRK model for quenches from a product state. We focus on the mutual information between two subregions after the quench, a quantity that provides a bound on the correlations functions~\cite{Wolf}.

Our analysis of the mutual information shows that even for very long-range interactions $\alpha<1$ only a small fraction of the mutual information violates locality. The largest build-up of mutual information occurs within a well defined `light cone'.

Surprisingly, the most important quantitative difference between the long and the short-range case is related to the decay, rather than the build-up, of mutual information. In the short-range case, mutual information is strongly peaked on the light cone itself, implying that information travels only as a localized wave packet. For long-range interactions, on the other hand, we find large mutual information at time-like separations as well. The longer persistence of mutual information here implies that also equilibration can be slowed down significantly.

We consider the following Hamiltonian on a lattice of length $L$:
\begin{eqnarray}
\label{eq:H_LRK}
\nonumber
H_\text{LRK} &=&-J \sum_{j=1}^{L}{ \left(c^\dagger_j c_{j+1} + c^\dagger_{j+1} c_j\right)} - \mu \sum_{j=1}^L {\left( c^\dagger_j c_j -\frac{1}{2}\right)} \\ \nonumber
&&+\Delta \sum_{j=1}^L{ \sum_{l=1}^{j-1}{ \left(\frac{c_j c_l + c^\dagger_l c^\dagger_j}{|l-j|^\alpha} \right)} }.
\end{eqnarray}

Here $c_j$ ($c^\dagger_j$) are the fermionic annihilation (creation) operators on the chain. The exponent $\alpha$ characterizes the range of the fermion pairing interactions, while the fermion hopping is only between nearest neighbours. 
We will set $J=\Delta= 1$ and send $L\rightarrow \infty$.

In the limit $\alpha \rightarrow \infty$, also the pairing term in (\ref{eq:H_LRK}) is between nearest neighbours only and Hamiltonian (\ref{eq:H_LRK}) can be mapped via a Jordan-Wigner transformation to the transverse-field Ising model \cite{jordanwigner}. In this situation, the phase diagram is symmetric for $\mu \leftrightarrow -\mu$ and the model has two critical points: $\mu=\pm 1$. They separate a ferromagnetic ($|\mu|<1$) and a paramagnetic phase ($|\mu|>1$) \cite{mussardo_book}.
For finite $\alpha$ the critical point at $\mu = 1$ persists, while the critical point at $\mu=-1$ disappears for $\alpha<1$. 

By defining $c_{k} = L^{-1/2} \sum_{l=1}^L{e^{-ik l} c_l}$, with the lattice momentum $k = 2\pi(n+1/2)/L$, we arrive at
\begin{equation}
\label{eq:H_LRK_k}
\nonumber
H = \sum_k \left[-(\cos{k}+\mu ) c^\dagger_{k} c_{k} + f^{(\alpha)}(k) \left( c_{k} c_{-k} + \text{h.c} \right) \right]
\end{equation}
where we define the functions $f^{(\alpha)}(k)= \sum_{u=1}^L {\sin{ku}/u^\alpha}$.
Via a Bogoliubov transformation $c_k=u_k \xi_k -iv_{-k} \xi^\dagger_{-k}$, the Hamiltonian (\ref{eq:H_LRK_k}) can be brought to diagonal form $H = E_0 + \sum_k{ \epsilon(k) \xi^\dagger_{k} \xi_{k}} $, with the quasiparticle dispersion
\begin{equation}
\label{eq:bogspec}
\epsilon(k) = \sqrt{ (\cos{k} +\mu)^2 + f^{(\alpha)}(k)^2},
\end{equation}
and $u_k = \cos(\theta_k/2)$ and $v_k=\sin(\theta_k/2)$, where $\tan\theta_k = -f^{(\alpha)}(k)/(\mu + \cos{k})$
\citep{LR_kitaev_Gorshkov}.

\begin{figure}[tbp]
\centering
\includegraphics[width=\columnwidth]{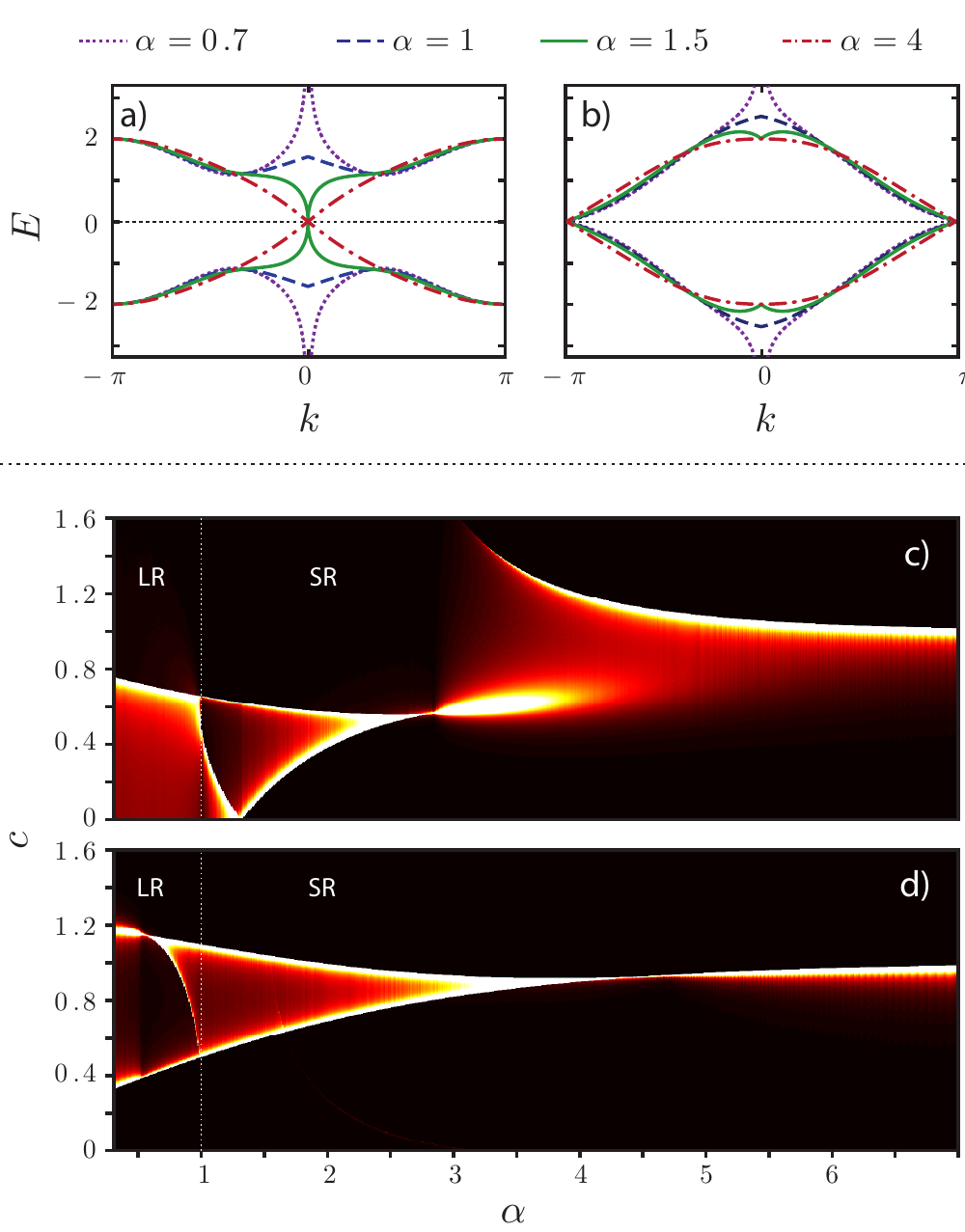}
\caption{(color online). Spectrum of LRK-model for $\mu=-1$ and $\mu=1$ are depicted in a) and b) respectively. c) and d) show the quasi-particle group velocity distribution for $\mu=-1$ and $\mu=1$ respectively. }
\label{fig:spectrum-velocity}
\end{figure}

The quasiparticle spectrum (\ref{eq:bogspec}) is gapped for all $\alpha>1$, expect for the critical lines $|\mu|=1$, where the gap closes. When $\alpha<1$ the dispersion diverges as $k^{\alpha-1}$ for $k\rightarrow0$. While this leads to a massive spectrum for $\mu=-1$, criticality is still preserved at $\mu = 1$, see Fig. \ref{fig:spectrum-velocity}~a) and \ref{fig:spectrum-velocity}~b). 

While gapped systems with finite-range interactions have an area law for entanglement entropy in the ground state \cite{AmicoRMP}, long-range interactions can lead to logarithmic corrections. In particular, a conformal field theory with effective central charge can be related to any system with $\alpha<1$. Likewise, the correlation function, which decays exponentially for short-range non-critical systems, has algebraic tails \cite{LR_kitaev_Gorshkov}, as was also observed in long-range Ising models \cite{LR_ising_entropy,LR_ising_ions,ising_dipolar,LR_ising_numerical}.

To study the propagation of information through the system, we will consider quenches in Hamiltonian (\ref{eq:H_LRK}) from $\mu = -\infty$, the noninteracting fermionic vacuum state, to $\mu = \pm 1$, and compute the subsequent time evolution (see the Supplemental Material).

 The quantities $\langle \alpha^\dagger_k \alpha_k \rangle = \sin^2(\theta_k/2)$ are conserved in time and must be included in the maximum-entropy ensemble at equilibrium \cite{jaynes,Rigol} (see the Supplemental Material).

For short-range interacting systems there is no correlation between two points at a distance $\Delta x$ up to a time $\Delta t=\Delta x/2c$, where $c$ is the Lieb-Robinson velocity. This is the minimal time it takes for an entangled particle hole pair to be shared between both points and is generally referred to as the light-cone effect. \cite{LR, Calabrese}.

Long-range interactions in turn can lead to an immediate correlation between distant points. In particular, we find for $\alpha<1$ that at large $r=|m-n|$ the correlations behave as $\langle c^\dagger_m c_n \rangle = i\langle c_m c_n \rangle = -\mathcal{F}^{(\alpha)}(r,t) \cos{[\eta^{(\alpha)}(r,t)]}$ The envelope has a power-law dependence in both time and distance: $\mathcal{F}^{(\alpha)}(r,t) = C^{(\alpha)}\cdot t^{\gamma} r^{-\chi}$, with $C^{(\alpha)}$ a constant. The scaling exponents are derived as (see the Supplemental Material)
\begin{equation}
\gamma = \frac{1}{2(2-\alpha)},\;\;\; \chi = \frac{3-\alpha}{2(2-\alpha)}.
\end{equation}

For the study of correlations  between spatially separated parts of the system, the quantum mutual information plays a central role, since it forms an upper bound on the correlation functions \cite{Wolf}.
It is defined as $I_{\mathcal{A},\mathcal{B}} = S_\mathcal{A}+S_\mathcal{B} - S_{\mathcal{A},\mathcal{B}}$ \cite{VedralRMP}, where $S_\mathcal{A} =- \text{tr} \rho_\mathcal{A} \log{\rho_\mathcal{A}}$ is the von Neumann entropy of the reduced density matrix on region $\mathcal{A}$ (see the Supplemental Material).

\begin{figure}[tbp]
\centering
\includegraphics[width=\columnwidth]{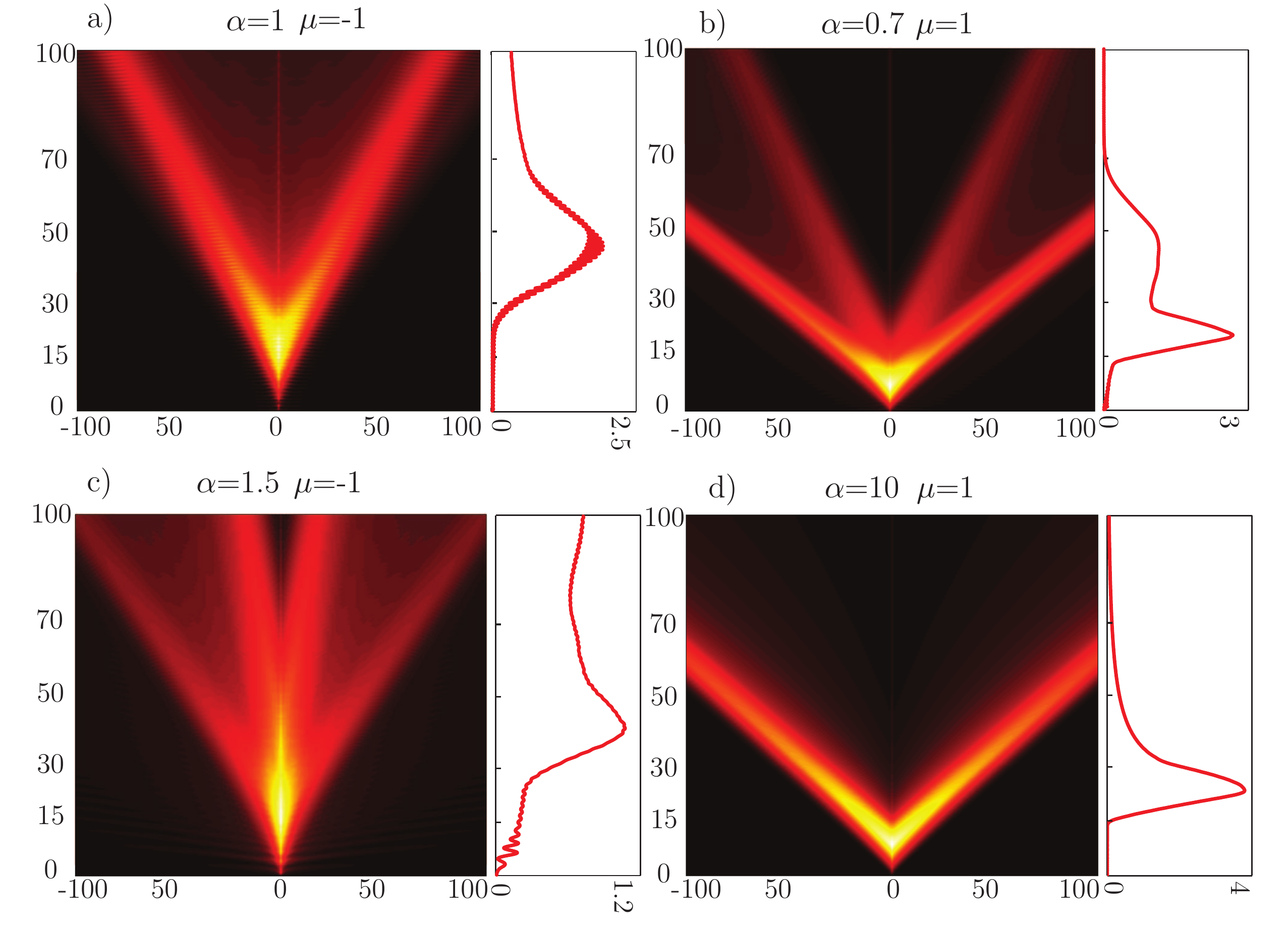}
\caption{(color online) Mutual information in the LRK chain after a global quantum quench as a function of subsystem separation and time. Subsystems are composed of 16 sites. All quenches start from the non-interacting ground state and go to a) $\alpha=1, \mu=-1$,b) $\alpha=0.7, \mu=1$, c) $\alpha=1.5, \mu=-1$ and d) $\alpha=10, \mu=1$. The side panels show the mutual information at a fixed separation of 30 sites.}
\label{fig:mutual}
\end{figure}

The Lieb-Robinson bound for short-range interacting systems is seen in the mutual information on Fig. \ref{fig:mutual}~d). At the time $t=\Delta x/2c$ there is a strong peak in the mutual information, which decays to zero again when the wave fronts have crossed each other. This is also the point where the joint system $\mathcal{A} \cup \mathcal{B}$ thermalizes. This image makes clear that information can propagate solely ballistically in a short-range interacting system \cite{LR,Calabrese}.

The situation changes drastically for long-range interactions. Immediately after the quench, information from one subsystem is shared with the other, regardless of their separation.  
Apart from the instantaneous rise of the mutual information, we also observe peaks in the mutual information, reminiscent of short-range systems. The increase of the mutual information due to the ballistic peaks is actually much larger than the nonlocal mutual information. 
The limiting case $\alpha=1,\; \mu=-1$ is quite peculiar, with almost perfect causality, despite the long-range nature of the interactions. 

We furthermore observe a strong dependence of the decay time of the mutual information on the interaction range.
For short range interactions, the decay is quite fast, as can be seen clearly in Fig.\ref{fig:mutual}~d), where the region inside the light-cone has very small mutual information. For long range interactions on the other hand, the decay can be much slower. The most striking examples are displayed in Figs.\ref{fig:mutual}~a)~and~c), where the mutual information inside the light cone remains significant at late times. We come back to this point below.

Since the LRK model is quadratic, the behavior of the quantum mutual information as a function of time and distance can be most easily understood from the quasi-particle group velocities $c = d\omega/dk$, which determine the rate at which
 information can propagate through the system.
For $\alpha\gg 1$, the quasi-particle velocity distribution $\mathcal{N}(c)$ is strongly peaked around its maximum velocity $c=\pm1$. Therefore information is carried ballistically across the system in strongly localized wave packets that travel at unit speed.


As was illustrated for short-range interacting systems, we can relate ballistic propagation to peaks in the velocity distribution $\mathcal{N}(c)$. In general the number of peaks and their location varies as a function of the Hamiltonian parameters $\alpha$ and $\mu$. Each peak corresponds to a wave front of information that travels at a finite speed through the system after the quench (see Fig. \ref{fig:spectrum-velocity}~c) and \ref{fig:spectrum-velocity}~d)).

For $\mu = 1$ there are three peaks or $\alpha<1$, which reduce to two at the crossover $\alpha=1$. For $\alpha\gg 1$ both peaks converge to $c = 1$, the group velocity of the short-range transverse Ising model.

The case $\mu=-1$ shows a much richer behaviour. Here there is only one peak at $\alpha<1$, which splits in two at $\alpha=1$.  At $\alpha\approx1.3$ one peak drops to zero velocity, corresponding to a ballistic wave packet with $c\approx 0$.
 Around $\alpha\approx3$, the peaks join and a new peak appears at high $c$. For $\alpha \gg 1$ both peaks converge again to $c = 1$.

For $\alpha<1$ the spectrum becomes singular around $k=0$, implying that the group velocity of modes with $k\approx 0 $ diverges. In contrast with short-range interacting systems, there no longer exists an upper bound for the group velocity. In the velocity distribution $\mathcal{N}(c)$, this manifests itself as long tails for $c\gg1$. Immediately after the quench, these ultrafast modes will cross large distances and correlate distant points in the chain.

The case $\mu=-1$ shows even more interesting behaviour. For $\alpha > 1$, we find that $c(k)|_{k\rightarrow 0}=\zeta(\alpha-1)=\sum_{k=1}^{\infty}{k^{1-\alpha}}$, which in the limit $\alpha\rightarrow \infty$ converges to $c(k)=1$, the ballistic velocity in the transverse-Ising model. Interestingly, when $1<\alpha <2$, the spectrum $\epsilon(k)$ is finite everywhere, but the group velocity $c$ still diverges around $k=0$. Therefore this case displays \textit{quasi long-range} behaviour in terms of ultrafast propagation of information.  
Most strikingly, the limiting case $\alpha=1$ has a finite spectrum and finite group velocity everywhere. A dominant quasiparticle velocity $c=1/2$ can be determined from the velocity distribution, thus recovering an effective light cone as in short-range interacting systems as shown in Fig.\ref{fig:mutual}~a).

Apart from the fast modes in the tail of the velocity distribution of long-range interacting systems, also very slow dynamics can be observed. Saddle points in $\epsilon(k)$ lead to occupations of modes with $c \approx 0$. This effect is visible for $\mu =-1$, where the spectrum has a minimum at non-zero $k$ for $\alpha\lesssim 1.3$. In the case $\alpha \approx 1.3$ this is most pronounced, as now there is even ballistic peak with $c\approx 0$.
At late times, these modes will spread information at a very slow rate and delay equilibration of the system. 
Note that the slowing down of the group velocity for $\mu=-1$ occurs in the intermediate interaction range and is not directly related to the range of the interactions. For decreasing $\alpha$, the equilibration of local observables speeds up again. The case $\mu=1$ in turn does not exhibit these ultraslow group velocities for any $\alpha$.



The slowing down of local equilibration is also reflected in the time evolution of the entanglement entropy itself. In general, the direct measurement of entanglement entropy is believed to be very hard, due to its strongly non-local nature, but recently promising methods were proposed \cite{Abanin,Daley} and implemented \cite{Islam}.

\begin{figure}[tbp]
\centering
\includegraphics[width=\columnwidth]{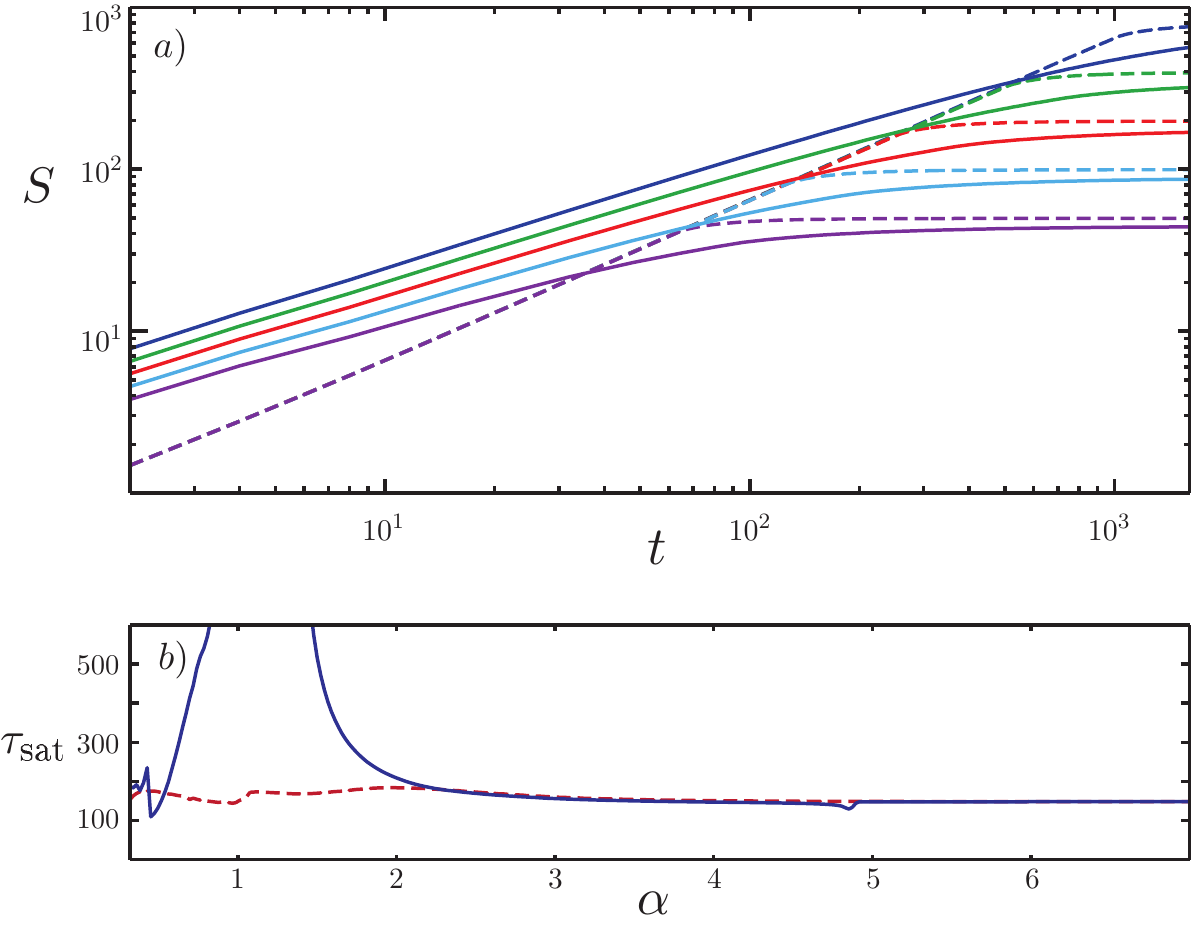}
\caption{(color online) a) Entanglement entropy as a function of time after a global quench in a LRK-chain. Results are shown for system subsystem sizes from 128 to 2048 sites. All quenches start from the non-interacting ground state to a value of $\mu=1$ and $\alpha=0.7$. For comparison, the dashed lines show the result for $\alpha=10$. b) The time $\tau_\text{sat}$ it takes to reach 95\% of the GGE entropy as a function of $\alpha$, for $\mu = -1$ (blue solid line) and $\mu = 1$ (red dashed line).  The subsystem size is $200$.}
\label{fig:entanglement}
\end{figure}

In Fig.\ref{fig:entanglement} we see in all cases that the entropy converges at late times to the GGE value, as is expected from the proof of \cite{barthel_dephasing} (see the Supplemental Material). However, the way in which the equilibration occurs is strongly dependent on the range of interactions. For short-range interactions ($\alpha \gg 1$) the effective light cone set up by the quasi-particles travelling at unit velocity, implies a linear growth of entanglement entropy before equilibration, as was also predicted for the corresponding CFT's \cite{Calabrese}. There is an abrupt saturation at $t_\text{sat}=L_\mathcal{A}/2$, the time it takes for the last particle hole pair, coming from the middle of the subsystem, to leave.
The curves for different $L_\mathcal{A}$ coincide up to the saturation time $t_\text{sat}$, as shown in Fig. \ref{fig:entanglement}.

For long-range interactions the initial entropy increase is faster than linear because of the ultrafast propagating modes. In particular, we find a power-law growth $S_\mathcal{A} \sim t^\beta$, with $\beta<1$. Surprisingly, this in contrast with previous results on the LRK model~\cite{Vodola}, on long-range spin chains~\cite{Schachenmayer} and coupled harmonic oscillators~\cite{nedzha}, where the initial entropy growth was found to be logarithmic. The time it takes to thermalize is prolonged by the slow modes in the velocity distribution (see Fig. \ref{fig:spectrum-velocity} c)). The transition to an equilibrated state is smooth and there is no clearly distinguishable saturation time.

In conclusion, it is possible to understand the main features of the information propagation and equilibration of local observables in long-range Kitaev chains in terms of its quasi-particle dispersion. Long-range ($\alpha<1$), as well as quasi long-range interactions ($\mu = -1$ and $1<\alpha<2$), give rise to an immediate increase of the mutual information after a quench from the noninteracting ground state, with the exception of the case $\mu=-1, \alpha=1$. 
For $\mu=-1$ and around $\alpha=1.3$, we also find large mutual information at time-like separation and slow equilibration to the generalized Gibbs ensemble, due to a vanishing of the dominant group velocity. It is an intriguing question whether a connection between long-range interactions and slow thermalization also exists for generic interacting quantum systems.

Shortly before the completion of this work, we became aware of similar ongoing work by Andrew Daley and Fabian Essler.

\acknowledgments
MVR gratefully acknowledges support in the form of a Ph. D. fellowship
of the Research Foundation - Flanders (FWO). MW acknowledges financial support from the FWO-Odysseus program. D.S. acknowledges support of the FWO as post-doctoral fellow of the Research Foundation - Flanders.

\bibliography{refs}

\end{document}



\title{Supplementary material to `Information propagation and equilibration in long-range Kitaev chains'}
\author{Mathias Van Regemortel}
\email{Mathias.VanRegemortel@uantwerpen.be}
\affiliation{TQC, Universiteit Antwerpen, B-2610 Antwerpen, Belgium}
\author{Dries Sels}
\affiliation{TQC, Universiteit Antwerpen, B-2610 Antwerpen, Belgium}
\affiliation{Department of Physics, Boston University, Boston, MA 02215, USA}
\author{Michiel Wouters}
\affiliation{TQC, Universiteit Antwerpen, B-2610 Antwerpen, Belgium}
\date{\today}

\date{\today}

\begin{abstract}

\end{abstract}

\pacs{}

\maketitle

\section{Construction of the density matrix}

We study systems with Hamiltonians of the general quadratic form 
\begin{equation}
\label{eq:Hquad}
H_{F,B} = \sum_{ij}{\left[c_i^\dagger V_{ij} c_j + \frac{1}{2}\left(c_i^\dagger W_{ij} c_j^\dagger+\text{h.c.} \right)\right]},
\end{equation}
where the $c^\dagger_i$ ($c_i$) denote fermionic creation (annihilation) operators on a lattice. 

The two-point normal and anomalous correlation functions, defined as $G^{(n)}_{ij} = \langle c^\dagger_i c_j \rangle$ and $G^{(a)}_{ij} = \langle c_i c_j \rangle$ respectively, allow for the reconstruction of the reduced density matrix of a subsystem $\mathcal{A}$, consisting of a set of $N_\mathcal{A}$ lattice sites.

A new set of operators $\gamma_q = \sum_{i\in \mathcal{A}}{ U_{qi} c_i + V_{qi} c^\dagger_i}$ exists, such that
\begin{equation}
\rho_\mathcal{A} = \prod_q{\frac{1}{n^{(\gamma)}_q}}	 \exp{\left(- \sum_q{\Omega_q \gamma^\dagger_q \gamma_q } \right)},
\end{equation} 
with $n^{(\gamma)}_q = \langle \gamma^\dagger_q \gamma_q \rangle = 1/(1+e^{-\Omega_q})$ the density of $\gamma_q$-modes.

The transformation matrices $U$ and $V$ and the pseudo-energies $\Omega_q$ are obtained by solving $P\mathcal{G}^{(+)}_\mathcal{A}= \Lambda_\mathcal{A} Q$ and $Q\mathcal{G}^{(-)}_\mathcal{A} = \Lambda_\mathcal{A} P$, where we defined
\begin{equation}
\mathcal{G}^{(\pm)}_\mathcal{A} = \matrx{2G^{(n)}_\mathcal{A}- \mathds{1}}{\pm 2G^{(m)}_\mathcal{A}}{{\pm 2G^{(m)}_\mathcal{A}}^\ast}{2G^{(n)}_\mathcal{A}- \mathds{1}},
\;\;\;
\Lambda_\mathcal{A} = \matrx{\Xi}{0}{0}{\Xi},
\end{equation}
with $\Xi_{qp} = \delta_{qp}\left(2 n^{(\gamma)}_q-1\right)$, and the unitary matrices
\begin{equation}
P = \matrx{U}{V}{V^\ast}{U^\ast},\;\;\; Q = \matrx{U}{-V}{-V^\ast}{U^\ast}.
\end{equation}
The entanglement entropy of $\mathcal{A}$ with its environment can now be evaluated as $S_\mathcal{A} = \sum_q{ h(n^{(\gamma)}_q)}$, with the entropy per mode
\begin{equation}
h(x) = -x\log{x}-(1-x)\log(1-x)
\end{equation}

\section{Relaxation to GGE}
\begin{figure}[tbp]
\centering
\includegraphics[width=\columnwidth]{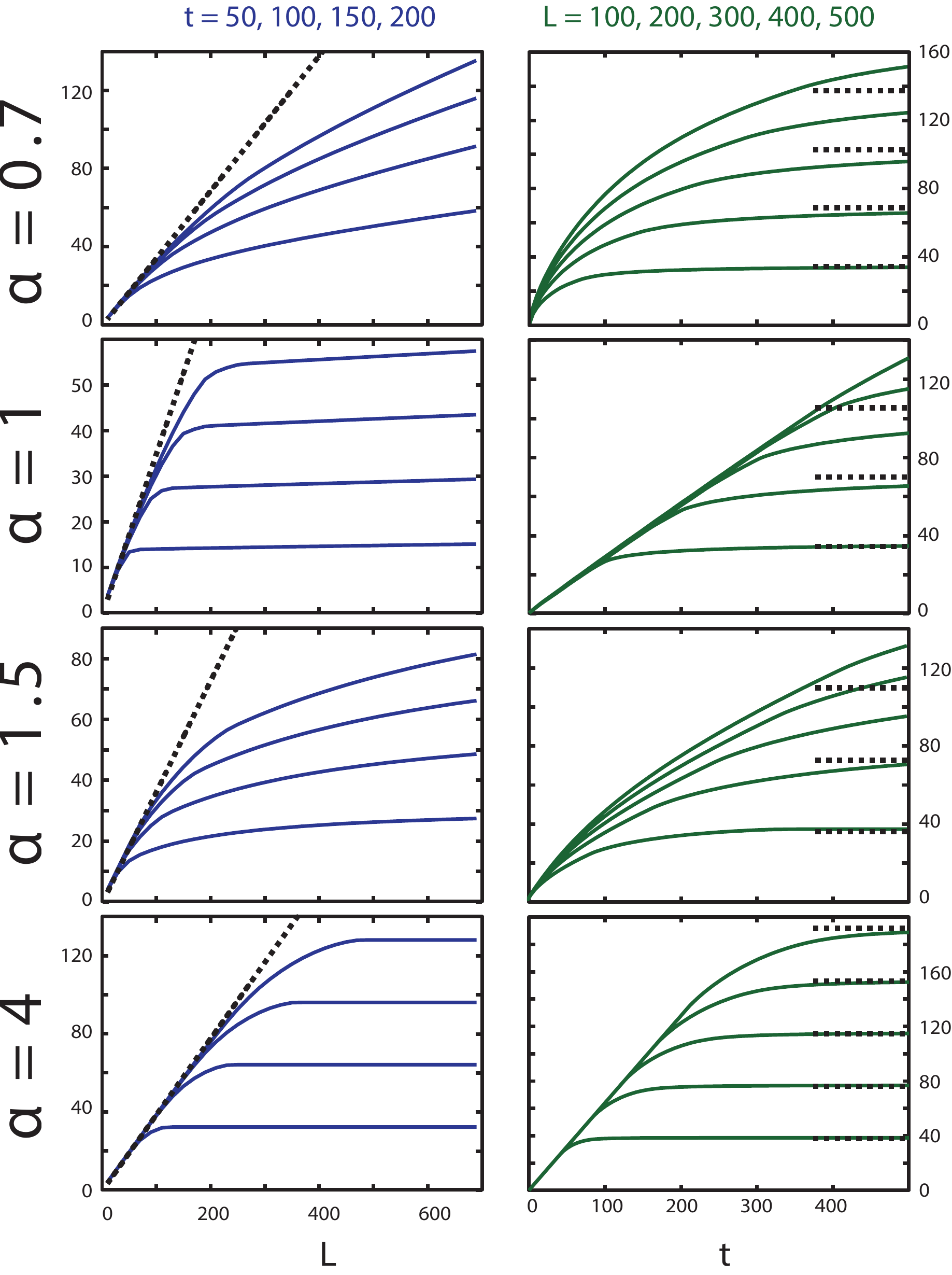}
\caption{The scaling of entanglement entropy as a function of subsystem size (left) and time (right) for different values of $\alpha$. The black dotted line marks the GGE result.}
\label{fig:entropies}
\end{figure}
The generalized Gibbs ensemble is the maximum-entropy ensemble at equilibrium \cite{Rigol}. It is constructed by inclusion of all conservation laws:
\begin{equation}
\label{eq:GGE}
\text{tr} \rho_\text{GGE} = \mathcal{Z}^{-1} \exp{-\left(\sum_k \lambda_k n_k \right)},
\end{equation}
with $n_k = v_k^2$, given in the main text, and $\lambda_k = \log{(u_k^2/v_k^2)}$, as set by the initial state.

The matrix constructed on $\mathcal{A}$, of which the eigenvalues $\mu_q = (2n^{(\gamma)}_q-1)^2$ yield  $S_\mathcal{A}$, reads
\begin{equation}
\label{eq:corrgamma}
\Gamma_\mathcal{A}= \mathcal{G}^{(+)}_\mathcal{A}\times\mathcal{G}^{(-)}_\mathcal{A} = \matrx{\Gamma_\mathcal{A}^{(d)}}{\Gamma_\mathcal{A}^{(a)}}{{\Gamma_\mathcal{A}^{(a)}}^\ast}{{\Gamma_\mathcal{A}^{(d)}}^\ast}.
\end{equation}
We find that $\langle c^\dagger_k c_{l} \rangle  = n^{(c)}\delta_{kl}$ and $\langle c_k c_{l} \rangle  = m^{(c)}\delta_{kl}$, with the density and anomalous correlation
\begin{eqnarray}
\label{eq:corr_k}
\nonumber
n_k^{(c)} &=&   \sin^2{\theta_k}\sin^2{\epsilon_k t}, \\\nonumber
m_k^{(c)} &=&  \sin{\theta_k} \left(1-\cos^2{\big(\frac{\theta_k}{2}\big)} e^{-i2\epsilon_k t} + \sin^2{\left(\frac{\theta_k}{2}\right)}  e^{i2\epsilon_k t} \right).
\end{eqnarray}
Using these, (\ref{eq:corrgamma}) can be written as:
\begin{eqnarray}
\label{eq:analytic_expr}
\nonumber
\Gamma_{mn}^{(d)} &=& \delta_{mn} -  \frac{4}{L}\sum_k {e^{-ik(m-n}n_k^{(c)}}\\ \nonumber
&&+ \frac{4}{L^2} \sum_{kl}{ e^{-ik m}e^{il n}\;W^\mathcal{A}_{kl}\;\big(n_k^{(c)} n_l^{(c)}+m_k^{(c)} m_l^{(c)}\big)} ,\\\nonumber
\Gamma_{mn}^{(a)}&=&\frac{4}{L^2} \sum_{kl}{ e^{-ikm}e^{ikn}\;W^\mathcal{A}_{kl}\;\big(m_k^{(c)} n_{l}^{(c)}-n_k^{(c)} m_{l}^{(c)}\big)},\\
&&
\end{eqnarray}
with $W^\mathcal{A}_{kl} = \sum_{n \in \mathcal{A}}e^{-i(k-l) n} $.

It is now easy to verify that the long-time limits of matrices (\ref{eq:analytic_expr}), with $m,n$ on a finite subsystem $\mathcal{A}$, are equivalent to their time averages 
\begin{equation}
\label{eq:corr_lim}
\overline{\Gamma}_{mn}^{(d)} = \sum_\kk {e^{-ik(m-n)}\left( 2v_k^2 -1\right)^2}\;, \;\;\;\;\;  \overline{\Gamma}_{mn}^{(a)} = 0,
\end{equation}
leading indeed to construction (\ref{eq:GGE}), the correct ensemble at equilibrium.

Fig. \ref{fig:entropies} shows the scaling of the entanglement entropy with system size (left panels) and time (right panels) for various values of the interaction range. It is clear that for longer interaction ranges (except for the special case $\alpha=1$), the thermalisation slows down, with a very slow approach to the GGE (dotted lines) for large subsystems.

\section{Derivation of the correlation function}
\begin{figure}[h!]
\centering
\includegraphics[width=\columnwidth]{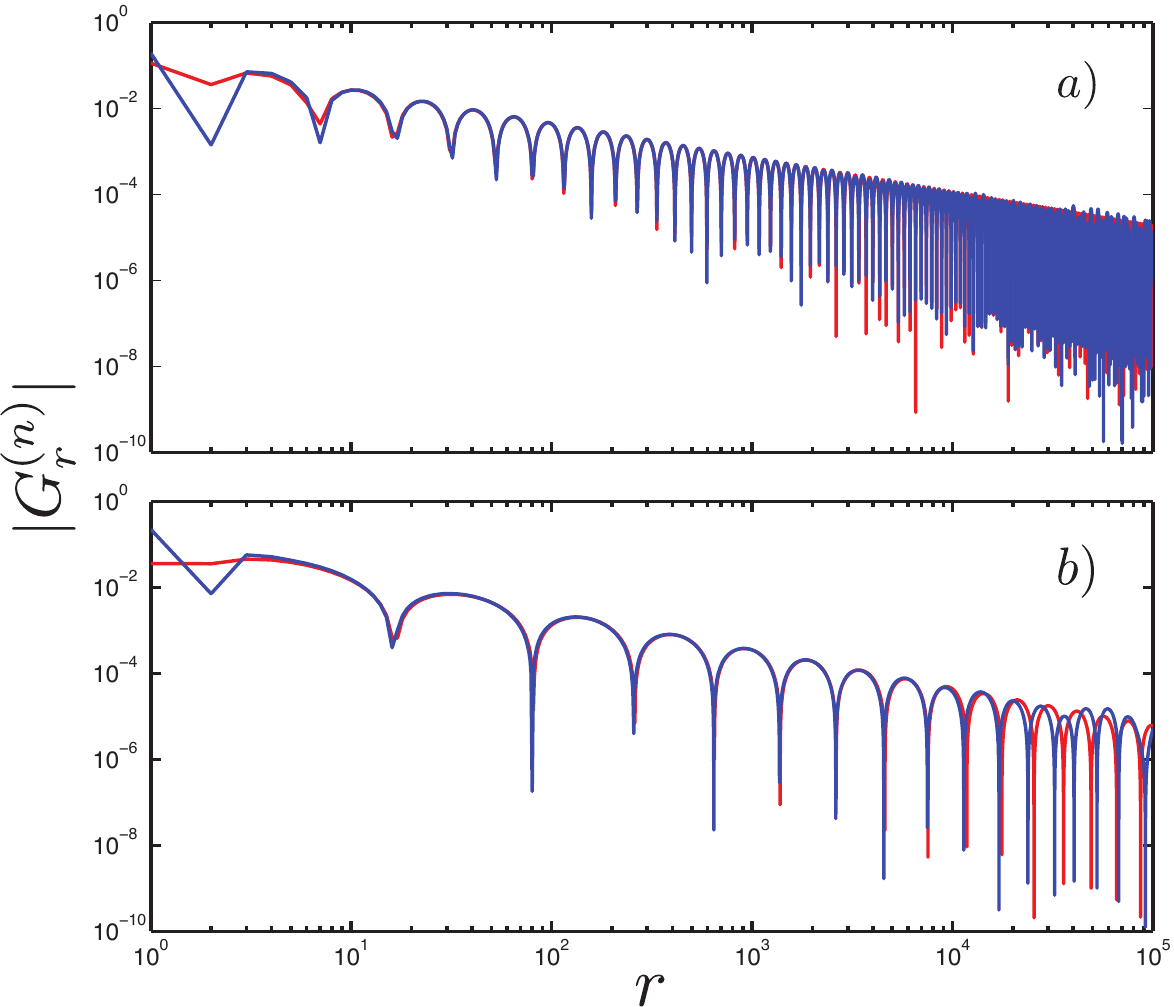}
\caption{The correlation function $G^{(n)}_r= \langle c_j^\dagger c_{j+r}\rangle$ (blue) and the approximation (red) from (\ref{eq:approx}) for $\alpha=0.3$ (a) and $\alpha = 0.7$ (b), both with $\mu=-1$.}
\label{fig:corr}
\end{figure}
The decay of the correlation functions at large distances can be evaluated for $\alpha <1$. The spectrum in the vicinity of the divergence can be approximated as $\epsilon_k\approx \xi^{(\alpha)} k^{\alpha-1}$, with $\xi^{(\alpha)} = \cos{(\pi \alpha/2)} \Gamma(1-\alpha)$ (see the Supplemental Material of \cite{Vodola}). Only modes close to the divergence will contribute at large distances.

Furthermore we have that $\theta_k \approx -\pi/2$ around the divergence, such that at large $r=|m-n|$,
\begin{equation}
\langle c^\dagger_m c_n \rangle = i\langle c_m c_n \rangle \approx -\frac{1}{4\pi} \Re \left(\int_{-\pi}^{\pi} {e^{-i(kr+2\epsilon_k t)}dk}\right)
\end{equation}
The integral can be evaluated with a static-phase approximation and yields expression (2) from the main text, with
\begin{eqnarray}
\label{eq:approx}
\nonumber
F(r,t)&=&\frac{1}{2\sqrt{2\pi(2-\alpha)}}\Big(2(1-\alpha) \xi^{(\alpha)}t \Big)^\gamma r^{-\chi},\\
\eta(r,t)&=& \frac{\pi}{4}+(2-\alpha)\Big(2\xi^{(\alpha)}t \Big)^{2\gamma}\Big((1-\alpha)r\Big)^{1-2\gamma}.
\end{eqnarray}
See Fig. \ref{fig:corr} for a comparison between the correlation function and this approximation.

%
%
\bibliography{refs}